\documentclass{elsart}

\usepackage{graphicx}
\usepackage{amssymb}

\begin{document}
\begin{frontmatter}
\title{Energy Estimators and Calculation of Energy Expectation Values in the Path Integral Formalism\thanksref{MNTR}}
\thanks[MNTR]{Supported by the Ministry of
Science and Environmental Protection of the Republic of Serbia through project No. 141035.}

\author{J. Gruji\' c\corauthref{jelenagr}},
\ead{jelenagr@phy.bg.ac.yu} \corauth[jelenagr]{Corresponding author.}
\author{A. Bogojevi\' c}, \and
\author{A. Bala\v z}
\address{Scientific Computing Laboratory, Institute of Physics\\
P. O. Box 57, 11001 Belgrade, Serbia}

\begin{abstract}
A recently developed method, introduced in \textit{Phys. Rev. Lett.} 94 (2005) 180403, \textit{Phys. Rev.} \textbf{B} 72 (2005) 064302, \textit{Phys. Lett.} \textbf{A} 344 (2005) 84, systematically improved the convergence of generic path integrals for transition amplitudes. This was achieved by analytically constructing a hierarchy of $N$-fold discretized effective actions $S^{(p)}_N$ labeled by a whole number $p$ and starting at $p=1$ from the naively discretized action in the mid-point prescription. The derivation guaranteed that the level $p$ effective actions lead to discretized transition amplitudes differing from the continuum limit by a term of order $1/N^p$. Here we extend the applicability of the above method to the calculation of energy expectation values. This is done by constructing analytical expressions for energy estimators of a general theory for each level $p$. As a result of this energy expectation values converge to the continuum as $1/N^p$. Finally, we perform a series of Monte Carlo simulations of several models, show explicitly the derived increase in convergence, and the ensuing speedup in numerical calculation of energy expectation values of many orders of magnitude.
\end{abstract}

\end{frontmatter}

\section{Introduction}
\label{sec:intro}

The path integral formalism first introduced by Feynman \cite{feynmanhibbs,feynman} represents a rich and flexible general mathematical setting for dealing with quantum and statistical theories. The most obvious success of path integrals has been the ease with which they have allowed us to extend the quantization procedure to ever more complicated systems. On a more practical note, the formalism has been extremely useful for handling symmetries, deriving non-perturbative results and connections between different theories \cite{itzyksonzuber,coleman}, and as a catalysts for the exchange of key ideas between different areas of physics, most notably high energy and condensed matter physics \cite{itzyksondrouffe,parisi}. Today, analytical and numerical approaches to path integrals \cite{barkerhenderson,pollockceperley,ceperley,kleinert} play important roles not only in physics but also in chemistry, materials science, mathematics and modern finance.

Further development of the path integral method is constrained by the small number of solvable models, as well as by our rather limited knowledge of their precise mathematical properties. In an attempt to fill this void a recent series of papers \cite{prl,prb} has investigated the dynamical implications of the property of stochastic self-similarity of path integrals by studying the relation between path integral discretizations of different coarseness. This has resulted in a systematic analytical construction of a hierarchy of $N$-fold discretized effective actions $S^{(p)}_N$ labeled by a whole number $p$ and built up from the naively discretized action in the mid-point prescription (corresponding to $p=1$). It was shown that the level $p$ effective actions lead to discretized transition amplitudes differing from the continuum limit by a term of order $1/N^p$. These analytical results in fact represent an explicit derivation of the generalization of Euler's summation formula to path integrals \cite{euler}.

From a numerical stand point the new method has lead to a many order of magnitude speedup of path integral simulations by making it possible to get precise results using small values of $N$. The substantial numerical speedup is the direct result of the new analytical input that makes it possible to trade high $N$ (high cost in computing time) for low $N$ and high $p$. For example, for $p=9$ the overall speedup in convergence of the algorithm is typically eight orders of magnitude \cite{prb} over the defining algorithm, or five orders of magnitude when compared to previous state-of-the-art algorithms \cite{takahashiimada,libroughton,jangetal} based on the generalized Trotter formula \cite{deraedt2}, or other short-time approximation schemes \cite{predescu}.

In this paper we extend the applicability of the above method from transition amplitudes to the calculation of expectation values. In particular, we focus on energy expectation values. It is well known that the efficient calculation of energy expectation values necessitates the use of the optimal energy estimators \cite{barker,creutzfreedman,hermanetal,parrinellorahman}. An excellent review of different energy estimators is given in \cite{jankesauer}. In order to increase convergence of energy expectation values to the continuum limit it is necessary to understand how the estimators need to change as we move through the hierarchy of effective actions. In fact, if estimator and effective action are not synchronized then the increase in convergence for transition amplitudes does not translate into a corresponding increase in convergence for energy expectation values. The central result of this paper is the explicit analytical construction of optimal energy estimators for each hierarchy level $p$. We show that energy expectation values calculated using $p$ level effective actions and the associated energy estimators derived here converge to the correct continuum value as $1/N^p$, i.e. lead to the same increase in convergence (and algorithm speedup) previously seen for transition amplitudes. We have carried out a series of Monte Carlo simulations of several models (anharmonic oscillator, P\"oschl-Teller potential, Morse potential) and have shown explicitly that the increase in convergence (and the ensuing speedup in numerical calculation of energy expectation values) agrees with the analytical derivations. The figures presented in the paper illustrate the results of these simulations for the case of an anharmonic oscillator with quartic coupling.

Section~\ref{sec:estimators} starts with a brief overview of notation and an introduction to different energy estimators. It then develops the procedure for determining the estimator associated to each effective action at hierarchy level $p$. Section~\ref{sec:numerical} presents the results of our Monte Carlo simulations. In the Appendix we give the explicit expressions for the energy estimators for $p\le 6$. The results for $p\le 9$ and the codes used can be found on our web site \cite{scl}.

\section{Energy estimators}
\label{sec:estimators}

In the functional formalism the transition amplitude (in Euclidean time) $A(a,b;T)=\langle b|e^{-T\hat H}|a\rangle$ is given in terms of a path integral which is simply the $N\to\infty$ limit of the $(N-1)$- fold integral expression
\begin{equation}
\label{amplitudeN} A_N(a,b;T)=\left(\frac{1}{2\pi\epsilon_N}\right)^{\frac{N}{2}}\int dq_1\cdots dq_{N-1}\,e^{-S_N}\ .
\end{equation}
The Euclidean time interval $[0,T]$ has been subdivided into $N$ equal time steps of length $\epsilon_N=T/N$, with $q_0=a$ and $q_N=b$. The integrand is given in terms of the naively discretized action $S_N$. For actions of the form
\begin{equation}
\label{action}
S=\int_0^Tdt\,\left(\frac{1}{2}\, \dot q^2+V(q)\right)\ ,
\end{equation}
the naively discretized action (in the mid-point ordering prescription) equals
\begin{equation}
\label{actionN}
S_N=\sum_{n=0}^{N-1}\left(\frac{\delta_n^2}{2\epsilon_N}+\epsilon_NV(\bar q_n)\right)\ ,
\end{equation}
where $\delta_n=q_{n+1}-q_n$, and $\bar q_n=\frac{1}{2}(q_{n+1}+q_n)$. Note that we use units in which $\hbar$ and particle mass have been set to unity.

As can be seen, the very definition of path integrals makes necessary the transition from continuum to the discretized theory. This discretization, however, is far from unique. We are free to introduce additional terms that explicitly vanish in the continuum limit. Although such additional terms do not change the continuum physics, they do affect the speed of convergence to that continuum limit. A recent series of papers \cite{prl,prb} has studied the relation between discretizations of different coarseness and has analytically constructed a hierarchy of effective actions $S^{(p)}_N$. The starting member of the hierarchy corresponds to $p=1$ and is given by Eq.~(\ref{actionN}), the naively discretized action in the mid-point prescription. The $p$-level effective action leads to discretized amplitudes that converge to their continuum values as
\begin{equation}
A^{(p)}_N(a,b;T)=A(a,b;T)+\textrm{O}\left((T/N)^p\right)\ .
\end{equation}

Before extending the above outlined procedure to improve the convergence of energy expectation values, we briefly review the standard procedure for their calculation. The canonical partition function of statistical mechanics
\begin{equation}
Z(\beta)= \textrm{Tr} e^{-\beta \hat{H}}= \int dq\,A(q, q; \beta)\ ,
\end{equation}
is given in terms of diagonal transition amplitudes where the inverse temperature $\beta$ plays the role of propagation time $T$.
From the above it directly follows that the partition function may be written in terms of a path integral -- the continuum limit of the $N$-fold integral
\begin{equation}
\label{partitionfunction}
Z_N(\beta)= \left(\frac{1}{2\pi\epsilon_N}\right)^{\frac{N}{2}}\int dq_1\cdots dq_{N-1} dq_N\,e^{-S_N}\ ,
\end{equation}
where integration is over all periodic trajectories $q(t)=q(t+\beta)$. In the discretized theory this simply implies that $q_0=q_N$.

Path integral Monte Carlo simulations use two different types of energy estimators -- ``kinetic" and ``virial". Both types of estimators are well known and have been extensively studied in \cite{barker,creutzfreedman,hermanetal,parrinellorahman,jankesauer}. We start with the ``kinetic'' estimator which follows from straightforward differentiation of the partition function. The (thermal) expectation value of the energy $U(\beta)$ is simply the continuum limit of
\begin{equation}
U_N(\beta)= -\frac{\partial}{\partial\beta}\ln Z_N(\beta)\ .
\end{equation}
Note that $\beta=N\epsilon_N$, so that $\partial/\partial\beta=(1/N)\partial/\partial\epsilon_N$. Using this we finally find that
\begin{equation}
U_N(\beta)=\frac{1}{Z_N}\left(\frac{1}{2\pi\epsilon_N}\right)^\frac{N}{2}
\int dq_1\ldots q_N\,\left[\frac{1}{2\epsilon_N}+\frac{\partial S_N}{\partial \epsilon_N}\right] e^{-S_N}\ .
\end{equation}
Putting in the naively discretized action we recover the so-called ``kinetic'' estimator of the energy
\begin{equation}
E_\textrm{kinetic}=\frac{N}{2T}-\frac{1}{N}\sum_{n=0}^{N-1} \frac{\delta_n^2}{2\epsilon_N^2}
+ \frac{1}{N} \sum_{n=0}^{N-1}V(\bar q_n)\ .
\end{equation}

The last term on the right hand side is clearly the potential energy estimator. The second term is what one would naively take to be the estimator for kinetic energy. This term, however, diverges in the continuum limit. In fact, the role of the first term in the above expression (the one coming from the path integral measure) is precisely to cancel this divergence. The above estimator, taken as a whole, is well behaved in the continuum limit. However, the fact that it is given as a difference of two diverging terms implies that its variance diverges with $N$. It is advantageous, therefore, to find another estimator having the same mean value but with smaller variance. This second type of energy estimator is called ``virial" as it is based on the virial theorem
\begin{equation}
\left\langle \frac{\hat{p}^2}{2m}\right\rangle=\frac{1}{2} \left\langle \hat{x}V'(\hat{x}) \right\rangle\ .
\end{equation}
To derive this in the path integral formalism we rescale the coordinates in Eq.~(\ref{partitionfunction}) according to $q_n\to\mu q_n$. By setting
$\mu^2= \epsilon_N$ all the $\beta$ dependence is removed from the path integral measure, and we get
\begin{equation}
Z_N(\beta)=\left(\frac{1}{2\pi}\right)\int dq_1\ldots dq_N\,
\exp\left\{ -\sum_{n=0}^{N-1} \left[ \frac{\delta_n^2}{2} + \epsilon_N V(\epsilon_N^2\,\bar q_n)\right] \right\}\ .
\end{equation}
Differentiating with respect to $\epsilon_N$ we find the ``virial" estimator for the energy to be
\begin{equation}
E_\textrm{virial}=\frac{1}{2N}\sum_{n=0}^{N-1} q_n V'(q_n) + \frac{1}{N} \sum_{n=0}^{N-1} V(q_n)\ .
\end{equation}
As expected, the new estimator is a sum of terms that are finite in the continuum limit and so leads to smaller variance. The remaining work presented in this paper uses the ``virial" estimator given above and its $p$ level generalizations. From now on we drop the designation ``virial".

As we have seen, the estimator follows directly from the naive discretized action. If instead of $S_N$ we use some other effective action $S^{(p)}_N$ in the same hierarchy we directly determine the associated estimator at this level. For example, the $p=2$ level effective action equals \cite{prl,prb} \begin{equation}
S_N^{(2)}=\sum_{n=0}^{N-1}\left[\frac{\delta_N^2}{2\epsilon_N}+\epsilon_N V(\bar{q_n})+ \frac{\epsilon_N^2}{12}V'(\bar{q_n})+\frac{\delta_n^2}{24} V''(\bar{q_n})\right]\ .
\end{equation}
Using the above procedure we easily determine the associated estimator to be
\begin{equation}
E_{p=2}=\frac{1}{N}\sum_{n=0}^{N-1} \left[V + \frac{\bar q_n}{2} V'+\frac{\epsilon_N}{6}V''+\frac{\delta_n^2}{12}V''+\frac{\bar q_n\epsilon_N}{24} V'''+ \frac{\bar q_n\delta_n^2}{48}V'''\right]\ ,
\end{equation}
where $V$ is shorthand notation for $V(\bar q_n)$. For higher $p$ levels one proceeds in precisely the same way. The effective actions become more complex as we move up the hierarchy, so that the corresponding $p$ level estimators are most easily calculated using using a standard package for algebraic calculations such as MATHEMATICA. Explicit expressions for the energy estimators for $p\le 6$ are given in the Appendix. The results for $p\le 9$ as well as the Monte Carlo codes used in this paper can be found on our web site \cite{scl}.

The outlined procedure of matching a discretized action to its corresponding estimator has the property that amplitudes (and partition functions) have the same continuum limit as energy expectation values. As a result, by consistently using $p$ level effective actions and associated estimators we are guaranteed to have
\begin{equation}
\label{convergence}
U_N^{(p)}(\beta)=U(\beta)+\textrm{O}\left((\beta/N)^p\right)\ .
\end{equation}
In the following section we present explicit Monte Carlo simulations that verify this behavior.

\section{Numerical results}
\label{sec:numerical}

The numerical simulations presented in this section were done using Grid-adapted Monte Carlo code and were run on EGEE-II and SEE-GRID-2 infrastructure \cite{egee,seegrid}. As already indicated, we carried out a series of Monte Carlo simulations for several models, including anharmonic oscillator, modified P\"oschl-Teller potential, and Morse potential. We determined explicitly that the speedup in convergence of energy expectation values of a general model precisely conforms to the analytical result given in Eq.~(\ref{convergence}). The figures presented in this section give the results of simulations for the case of an anharmonic oscillator with quartic coupling using different levels $p$. The potential is thus
\begin{equation}
V(q)= \frac{1}{2}q^2+\frac{g}{4!}q^4\ .
\end{equation}
All presented results correspond to $g=24$ and $\beta=1$, however, the same kind of increase in convergence is seen for all values of coupling and inverse temperature (and, in fact, for all models numerically analyzed).

%------------------------------------------------------------------------------
\begin{figure}[!ht]
\centering
\includegraphics[width=13.cm]{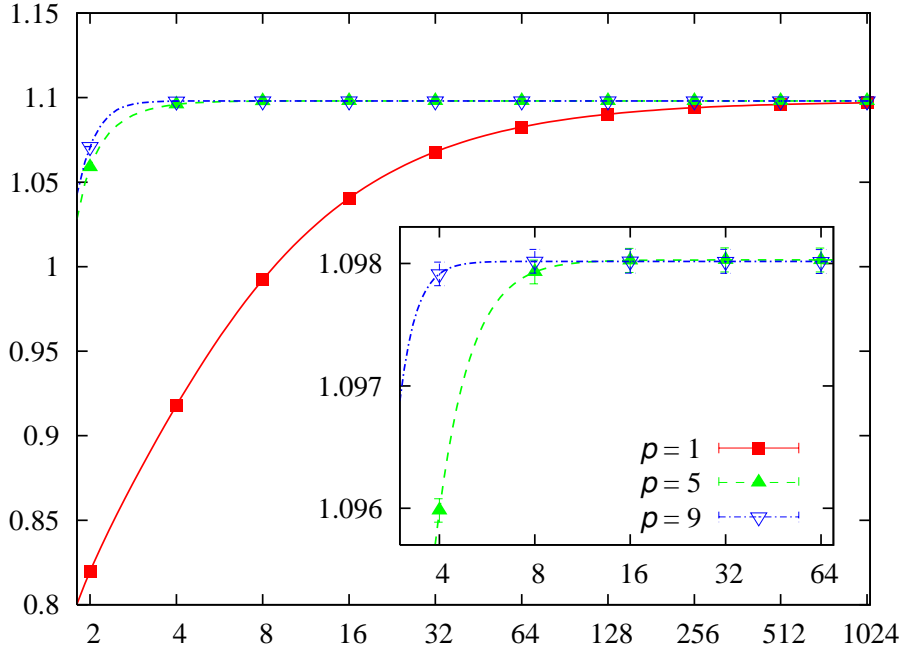}
\caption{\label{p159fit} Discretized energy expectation values $U_N^{(p)}$ as functions of $N$ for $p=1,5,9$ for the case of an anharmonic oscillator with quartic coupling $g=24$, time of propagation $\beta=1$, and $N_{MC}=10^9$. The lines represent appropriate $1/N$ polynomial fits to the data. The level $p$ curves have $1/N^p$ leading behavior.}
\end{figure}
%------------------------------------------------------------------------------

Fig.~\ref{p159fit} shows how discretized energy expectation values $U_N^{(p)}$ converge to the continuum for $p=1$ (naively discretized action and standard ``virial'' estimator), as well as for the effective actions and estimators at levels $p=5$ and $p=9$. From the figure we see that effective actions and estimators outperform the naively discretized action and standard estimator. The lines represent curve fits of the data to polynomials in $1/N$. We see that all $p$ levels have the same continuum limit (within the error). The leading behavior of the $p$ level curve fit is $1/N^p$. The inset plot gives a zoomed in view clearly showing that for $p=9$ one obtains excellent convergence (energy expectation value to five significant figures) for even  extremely coarse discretizations such as $N=3$.

\begin{figure}[!ht]
\centering
\includegraphics[width=13.cm]{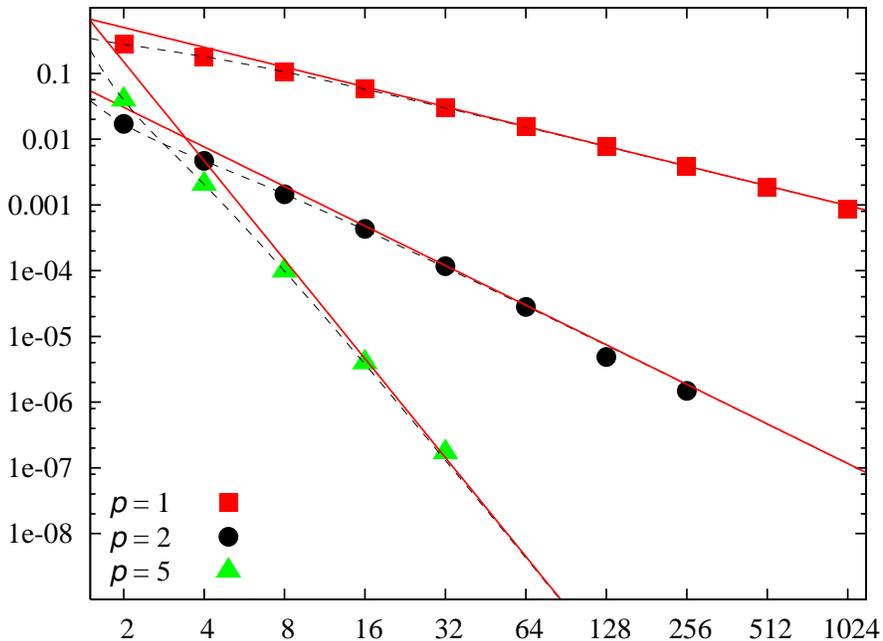}
\caption{\label{p125log} Log-log plot of the deviations from the continuum limit $|U_N^{(p)}-U|$ as functions of $N$ for $p=1,2$ and $5$ for an anharmonic oscillator with quartic coupling $g=24$, time of propagation $\beta=1$, and $N_{MC}=10^9$. Dashed lines correspond to appropriate $1/N$ polynomial fits to the data. Solid lines give the leading $1/N$ behavior showing that the level $p$ curve has $1/N^p$ leading behavior.}
\end{figure}

An even more clear-cut demonstration of the fact that numerical simulations conform to Eq.~(\ref{convergence}) can be seen in Fig \ref{p125log}. The plot display deviations from the continuum limit as a function of $N$. The dashed lines are polynomial data fits in $1/N$ starting from $1/N^p$, the leading terms $1/N^p$ are given as solid lines. The deviations from the continuum limit $|U_N^{(p)}-U|$ become extremely small for larger values of $N$ and $p$ and are
masked by statistical Monte Carlo errors. For this reason, one would need to use an even larger $N_{MC}$ in order to show the $p=9$ level deviation curve. Said another way, the $p=9$ result for $N=4$ is already well within the statistical error for $N_{MC}=10^{9}$.
%------------------------------------------------------------------------------
%------------------------------------------------------------------------------

The same behavior was seen for other values of parameters as well as for the two other models tested: particle in a modified P\" oschl-Teller potential  $V_{mPT}(q)=-\frac{1}{2}[\chi ^2\lambda(\lambda-1)/\cosh^2(\chi q)]$ for $\chi =0.5$ and $\lambda =15.5$, and particle in a Morse potential
$V_M(q)= C(e^{-2\kappa q}-2e^{-\kappa q})$ for $C=10$ and $\kappa= 2$.

\begin{figure}[!ht]
\centering
\includegraphics[width=13.cm]{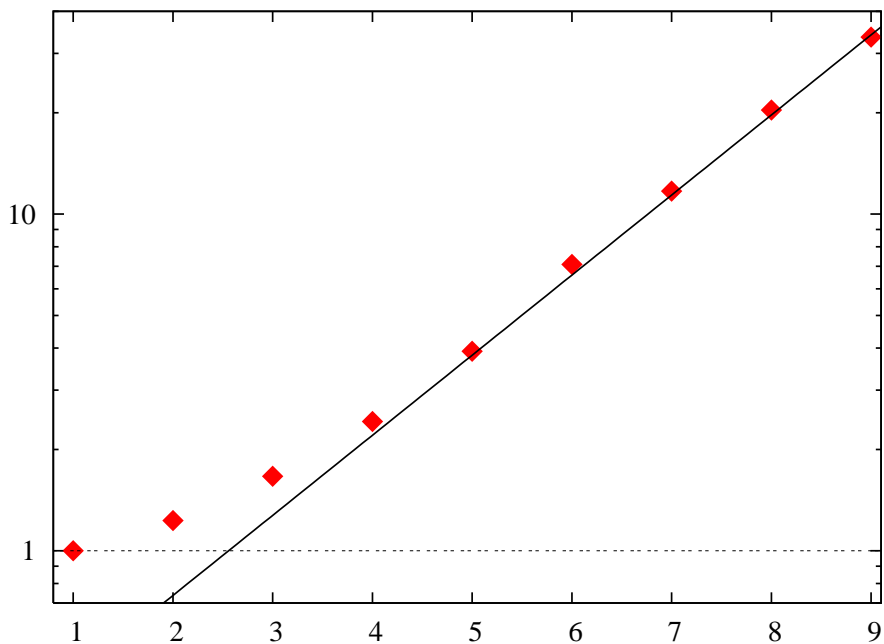}
\caption{\label{time} Relative increase in computation time that comes from the increased complexity of expression for higher $p$-level effective actions and energy estimators for fixed $N$. For $p\ge 5$ this ``complexity penalty" is well approximated by an exponential -- a direct consequence of the fact that $p$ level effective actions follow from the starting action via a $p$-fold recursive process.}
\end{figure}

As mentioned earlier, the effective actions and energy estimators for a give $N$ get more complex as we go to higher values of $p$. Consequently, there is a relative increase in computation time as we increase $p$. Fig.~\ref{time} shows this dependence as a function of $p$. We notice that for $p\ge 5$ the dependence is essentially exponential. Because of this increase in complexity, computations using $S^{(9)}_N$ are about 37 times slower than that with $S_N$.
On the other hand, increase of $p$ drastically improves convergence to the continuum limit, making it possible to obtain the same precision using much smaller values of $N$. Even for $p=9$ the gain in precision outweighs the increase in computation time arising from complexity by many orders of magnitude.
%------------------------------------------------------------------------------
%------------------------------------------------------------------------------

To conclude, we have extended the method for systematically speeding up path integral calculation introduced in \cite{prl,prb} to calculation of energy expectation values. We have shown that a consistent choice of effective action and estimator leads to the same form of speedup for expectation values that was previously seen for transition amplitudes, i.e. at $p$ level the discretized expectation values differ from the continuum as $1/N^p$.

\appendix

\section{Energy estimators to $p=6$ level}

In the following we use the short hand notation $q=\bar q_n$, $V=V(\bar q_n)$, $\delta^2 = \delta^2_n$, and $\epsilon =\epsilon_N = \beta/N$.

\begin{equation}
E_{p=1} = \frac{1}{N} \sum_{n=0}^{N-1} \left[ V +\frac{q\,V'}{2} \right].
\end{equation}

\begin{equation}
E_{p=2} = E_{p=1} +  \frac{1}{N}\sum_{n=0}^{N-1} \left[ \frac{\epsilon\, V''}{6} + \frac{\delta^2\, V''}{12} + \frac{q\, \epsilon\, V^{(3)}}{24} + \frac{ q\, \delta^2\, V^{(3)}}{48}\right]
\end{equation}

\begin{eqnarray}
&&E_{p=3}  =  E_{p=2} + \frac{1}{N} \sum_{n=0}^{N} \left[ - \frac{{\epsilon }^2\,{V'}^2}{8} \right.- \frac{q\,{\epsilon }^2\,V'\,V''}{24} + \frac{{\delta }^4\,V^{(4)}}{640} + \frac{{\delta }^2\,\epsilon \,V^{(4)}}{160}\nonumber\\
& & \,+ \frac{{\epsilon }^2\,V^{(4)}}{80} + \frac{q\,{\delta }^4\,V^{(5)}}{3840} + \frac{q\,{\delta }^2\,\epsilon \,V^{(5)}}{960}\left. +
  \frac{q\,{\epsilon }^2\,V^{(5)}}{480} \right]
\end{eqnarray}

\begin{eqnarray}
&&E_{p=4}  =  E_{p=3} + \frac{1}{N} \sum_{n=0}^{N}\left[
- \frac{ {\delta }^2\,{\epsilon }^2\,{V''}^2 }{360} - \frac{{\epsilon }^3\,{V''}^2}{90} - \frac{{\delta }^2\,{\epsilon }^2\,V'\,V^{(3)}}{120} \right. -   \frac{{\epsilon }^3\,V'\,V^{(3)}}{30} \nonumber\\
& & -\, \frac{q\,{\delta }^2\,{\epsilon }^2\,V''\,V^{(3)}}{576} - \frac{q\,{\epsilon }^3\,V''\,V^{(3)}}{144} -
  \frac{q\,{\delta }^2\,{\epsilon }^2\,V'\,V^{(4)}}{960} - \frac{q\,{\epsilon }^3\,V'\,V^{(4)}}{240}  \nonumber\\
& & +\, \frac{{\delta }^6\,V^{(6)}}{80640} +
  \frac{{\delta }^4\,\epsilon \,V^{(6)}}{13440} + \frac{{\delta }^2\,{\epsilon }^2\,V^{(6)}}{3360} + \frac{{\epsilon }^3\,V^{(6)}}{1680} +
  \frac{q\,{\delta }^6\,V^{(7)}}{645120} \nonumber\\
& & +\, \frac{q\,{\delta }^4\,\epsilon \,V^{(7)}}{107520} + \frac{q\,{\delta }^2\,{\epsilon }^2\,V^{(7)}}{26880} +  \left. \frac{q\,{\epsilon }^3\,V^{(7)}}{13440}
 \right]
\end{eqnarray}

\begin{eqnarray}
&&E_{p=5}  =  E_{p=4} + \frac{1}{N} \sum_{n=0}^{N}\left[
\frac{{\epsilon }^4\,{V'}^2\,V''}{48} \right. + \frac{q\,{\epsilon }^4\,V'\,{V''}^2}{240} + \frac{q\,{\epsilon }^4\,{V'}^2\,V^{(3)}}{480}  \nonumber\\
& & -\, \frac{5\,{\delta }^4\,{\epsilon }^2\,{V^{(3)}}^2}{32256}- \frac{5\,{\delta }^2\,{\epsilon }^3\,{V^{(3)}}^2}{4032} -
  \frac{23\,{\epsilon }^4\,{V^{(3)}}^2}{8064} - \frac{{\delta }^4\,{\epsilon }^2\,V''\,V^{(4)}}{8064}  \nonumber\\
& & -\,\frac{{\delta }^2\,{\epsilon }^3\,V''\,V^{(4)}}{1008} -
  \frac{{\epsilon }^4\,V''\,V^{(4)}}{336} - \frac{q\,{\delta }^4\,{\epsilon }^2\,V^{(3)}\,V^{(4)}}{23040} -
  \frac{q\,{\delta }^2\,{\epsilon }^3\,V^{(3)}\,V^{(4)}}{2880}  \nonumber\\
& & -\, \frac{q\,{\epsilon }^4\,V^{(3)}\,V^{(4)}}{1152} -
  \frac{{\delta }^4\,{\epsilon }^2\,V'\,V^{(5)}}{10752} - \frac{{\delta }^2\,{\epsilon }^3\,V'\,V^{(5)}}{1344} - \frac{{\epsilon }^4\,V'\,V^{(5)}}{448}  \nonumber\\
& & -\,  \frac{q\,{\delta }^4\,{\epsilon }^2\,V''\,V^{(5)}}{46080} - \frac{q\,{\delta }^2\,{\epsilon }^3\,V''\,V^{(5)}}{5760} -
  \frac{q\,{\epsilon }^4\,V''\,V^{(5)}}{1920} - \frac{q\,{\delta }^4\,{\epsilon }^2\,V'\,V^{(6)}}{107520} \nonumber\\
& & -\, \frac{q\,{\delta }^2\,{\epsilon }^3\,V'\,V^{(6)}}{13440} -  \frac{q\,{\epsilon }^4\,V'\,V^{(6)}}{4480} + \frac{{\delta }^8\,V^{(8)}}{18579456} +
  \frac{{\delta }^6\,\epsilon \,V^{(8)}}{2322432} + \frac{{\delta }^4\,{\epsilon }^2\,V^{(8)}}{387072} \nonumber\\
& & +\, \frac{{\delta }^2\,{\epsilon }^3\,V^{(8)}}{96768} +
  \frac{{\epsilon }^4\,V^{(8)}}{48384} + \frac{q\,{\delta }^8\,V^{(9)}}{185794560} + \frac{q\,{\delta }^6\,\epsilon \,V^{(9)}}{23224320} +
  \frac{q\,{\delta }^4\,{\epsilon }^2\,V^{(9)}}{3870720} \nonumber\\
& & +\, \frac{q\,{\delta }^2\,{\epsilon }^3\,V^{(9)}}{967680} + \left. \frac{q\,{\epsilon }^4\,V^{(9)}}{483840}
\right]
\end{eqnarray}

\begin{eqnarray}
&&E_{p=6}  =  E_{p=5}+ \frac{1}{N} \sum_{n=0}^{N}\left[ \frac{{\delta }^2\,{\epsilon }^4\,{V''}^3}{10080} \right. + \frac{{\epsilon }^5\,{V''}^3}{945} + \frac{{\delta }^2\,{\epsilon }^4\,V'\,V''\,V^{(3)}}{560} +
  \frac{29\,{\epsilon }^5\,V'\,V''\,V^{(3)}}{3360} \nonumber\\
& & +\, \frac{q\,{\delta }^2\,{\epsilon }^4\,{V''}^2\,V^{(3)}}{5760} +
  \frac{17\,q\,{\epsilon }^5\,{V''}^2\,V^{(3)}}{17280} + \frac{q\,{\delta }^2\,{\epsilon }^4\,V'\,{V^{(3)}}^2}{6720} +
  \frac{29\,q\,{\epsilon }^5\,V'\,{V^{(3)}}^2}{40320}\nonumber\\
& & +\, \frac{{\delta }^2\,{\epsilon }^4\,{V'}^2\,V^{(4)}}{2240} + \frac{3\,{\epsilon }^5\,{V'}^2\,V^{(4)}}{1120} +
  \frac{q\,{\delta }^2\,{\epsilon }^4\,V'\,V''\,V^{(4)}}{4480} + \frac{47\,q\,{\epsilon }^5\,V'\,V''\,V^{(4)}}{40320}\nonumber\\
& & -\, \frac{{\delta }^6\,{\epsilon }^2\,{V^{(4)}}^2}{691200} - \frac{{\delta }^4\,{\epsilon }^3\,{V^{(4)}}^2}{57600} -
  \frac{13\,{\delta }^2\,{\epsilon }^4\,{V^{(4)}}^2}{134400} - \frac{47\,{\epsilon }^5\,{V^{(4)}}^2}{201600} +
  \frac{q\,{\delta }^2\,{\epsilon }^4\,{V'}^2\,V^{(5)}}{26880}\nonumber\\
& & +\, \frac{q\,{\epsilon }^5\,{V'}^2\,V^{(5)}}{4480} -
  \frac{{\delta }^6\,{\epsilon }^2\,V^{(3)}\,V^{(5)}}{276480} - \frac{{\delta }^4\,{\epsilon }^3\,V^{(3)}\,V^{(5)}}{23040} -
  \frac{{\delta }^2\,{\epsilon }^4\,V^{(3)}\,V^{(5)}}{4480}\nonumber\\
& & -\, \frac{19\,{\epsilon }^5\,V^{(3)}\,V^{(5)}}{40320} -
  \frac{q\,{\delta }^6\,{\epsilon }^2\,V^{(4)}\,V^{(5)}}{1843200} - \frac{q\,{\delta }^4\,{\epsilon }^3\,V^{(4)}\,V^{(5)}}{153600} -
  \frac{q\,{\delta }^2\,{\epsilon }^4\,V^{(4)}\,V^{(5)}}{28800}\nonumber\\
& & -\, \frac{q\,{\epsilon }^5\,V^{(4)}\,V^{(5)}}{12800} -
  \frac{{\delta }^6\,{\epsilon }^2\,V''\,V^{(6)}}{967680} - \frac{{\delta }^4\,{\epsilon }^3\,V''\,V^{(6)}}{80640} -
  \frac{{\delta }^2\,{\epsilon }^4\,V''\,V^{(6)}}{13440} - \frac{{\epsilon }^5\,V''\,V^{(6)}}{5040}\nonumber\\
& & -\,
  \frac{q\,{\delta }^6\,{\epsilon }^2\,V^{(3)}\,V^{(6)}}{2580480} - \frac{q\,{\delta }^4\,{\epsilon }^3\,V^{(3)}\,V^{(6)}}{215040} -
  \frac{q\,{\delta }^2\,{\epsilon }^4\,V^{(3)}\,V^{(6)}}{40320} - \frac{q\,{\epsilon }^5\,V^{(3)}\,V^{(6)}}{17920}\nonumber\\
& & -\, \frac{{\delta }^6\,{\epsilon }^2\,V'\,V^{(7)}}{1935360} - \frac{{\delta }^4\,{\epsilon }^3\,V'\,V^{(7)}}{161280} -
  \frac{{\delta }^2\,{\epsilon }^4\,V'\,V^{(7)}}{26880} - \frac{{\epsilon }^5\,V'\,V^{(7)}}{10080} - \frac{q\,{\delta }^6\,{\epsilon }^2\,V''\,V^{(7)}}{7741440}\nonumber\\
& & -\, \frac{q\,{\delta }^4\,{\epsilon }^3\,V''\,V^{(7)}}{645120} - \frac{q\,{\delta }^2\,{\epsilon }^4\,V''\,V^{(7)}}{107520} -
  \frac{q\,{\epsilon }^5\,V''\,V^{(7)}}{40320} - \frac{q\,{\delta }^6\,{\epsilon }^2\,V'\,V^{(8)}}{23224320}\nonumber\\
& & -\, \frac{q\,{\delta }^4\,{\epsilon }^3\,V'\,V^{(8)}}{1935360} - \frac{q\,{\delta }^2\,{\epsilon }^4\,V'\,V^{(8)}}{322560} -
  \frac{q\,{\epsilon }^5\,V'\,V^{(8)}}{120960} + \frac{{\delta }^{10}\,V^{(10)}}{6812467200} + \frac{{\delta }^8\,\epsilon \,V^{(10)}}{681246720} \nonumber\\
& & +\,  \frac{{\delta }^6\,{\epsilon }^2\,V^{(10)}}{85155840} + \frac{{\delta }^4\,{\epsilon }^3\,V^{(10)}}{14192640} +
  \frac{{\delta }^2\,{\epsilon }^4\,V^{(10)}}{3548160} + \frac{{\epsilon }^5\,V^{(10)}}{1774080} + \frac{q\,{\delta }^{10}\,V^{(11)}}{81749606400}\nonumber\\
& & +\, \frac{q\,{\delta }^8\,\epsilon \,V^{(11)}}{8174960640} + \frac{q\,{\delta }^6\,{\epsilon }^2\,V^{(11)}}{1021870080} +
  \frac{q\,{\delta }^4\,{\epsilon }^3\,V^{(11)}}{170311680} +\nonumber\\
& & +\frac{q\,{\delta }^2\,{\epsilon }^4\,V^{(11)}}{42577920} + \left. \frac{q\,{\epsilon }^5\,V^{(11)}}{21288960} \right]
\end{eqnarray}

\end{document}